\documentclass[twocolumn]{article}  
\usepackage{times}  
\usepackage{helvet}  
\usepackage{courier}  
\usepackage[hyphens]{url}  
\usepackage{graphicx} 
\usepackage[numbers]{natbib}  
\usepackage{caption} 
\usepackage{algorithm}
\usepackage{algorithmic}

\usepackage{newfloat}
\usepackage{listings}
\DeclareCaptionStyle{ruled}{labelfont=normalfont,labelsep=colon,strut=off} 
\floatstyle{ruled}
\newfloat{listing}{tb}{lst}{}
\floatname{listing}{Listing}
\usepackage{authblk}
\usepackage{enumitem}
\usepackage{tikz}
\usepackage{pgfplots}
\pgfplotsset{compat=1.18}
\usepackage{pgfplotstable}
\usepackage{booktabs}
\usepgfplotslibrary{statistics, groupplots}
\usepackage{xcolor}
\definecolor{pptorange}{RGB}{237,125,49}
\definecolor{pptblue}{RGB}{68,114,196}
\definecolor{pptpurple}{RGB}{112,48,160}
\definecolor{pptred}{RGB}{192,0,0}
\definecolor{pptgreen}{RGB}{0, 97, 0} 
\definecolor{pptbrown}{RGB}{153, 102, 51}
\definecolor{pptdarkgrey}{RGB}{64, 64, 64}
\definecolor{pptdarkgreen}{RGB}{0, 112, 33}
\definecolor{pptteal}{RGB}{0,128,192}

\pgfplotsset{every axis/.append style={
    xlabel={$x$},          
    ylabel={$y$},          
    label style={font=\sffamily},
    tick label style={font=\sffamily\scriptsize},
    xticklabel style = {font=\sffamily\scriptsize},
    title style = {font=\normalsize\sffamily},
    ylabel near ticks,
    y label style={font=\sffamily\small},
    xlabel near ticks,
    x label style={font=\sffamily\small},
    },
    }

\newcommand*{\eg}{{\em e.g.}}

\newcommand*{\ie}{{\em i.e.}}

\author[1]{Adnan Hoq}
\author[1]{Matthew Facciani}
\author[1]{Tim Weninger}

\affil[1]{University of Notre Dame \\
         \{ahoq, mfacciani, tweninger\}@nd.edu
         }

\date{} 
\title{AI Credibility Signals Outrank Institutions and Engagement in Shaping News Perception on Social Media}

\begin{document}

\maketitle

\begin{abstract}
AI-generated content is rapidly becoming a salient component of online information ecosystems, yet its influence on public trust and epistemic judgments remains poorly understood. We present a large-scale mixed-design experiment (N = 1,000) investigating how AI-generated credibility scores affect user perception of political news. Our results reveal that AI feedback significantly moderates partisan bias and institutional distrust, surpassing traditional engagement signals such as likes and shares. These findings demonstrate the persuasive power of generative AI and suggest a need for design strategies that balance epistemic influence with user autonomy.
\end{abstract}

\section{Introduction}

Social media has become a dominant source of news, reshaping how people encounter and evaluate information~\cite{walker2021news, fletcher2018people, bakshy2015exposure}. Unlike traditional media governed by editorial norms~\cite{tandoc2018news, waisbord2013reinventing}, social platforms rely on dynamic, often opaque, sociotechnical infrastructures that mediate content through engagement signals (\eg, likes, shares, comments) as well as increasingly through generative AI~\cite{hermida2012share, metzger2010social, lustig2016algorithmic}. These cues act as heuristic markers of credibility~\cite{muchnik2013social, messing2014selective}, shaping public discourse in ways that are difficult for users to fully trace or interrogate~\cite{scheufele2007framing, molden2014understanding}.

\begin{figure}
    \centering
    \includegraphics[width=0.99\linewidth]{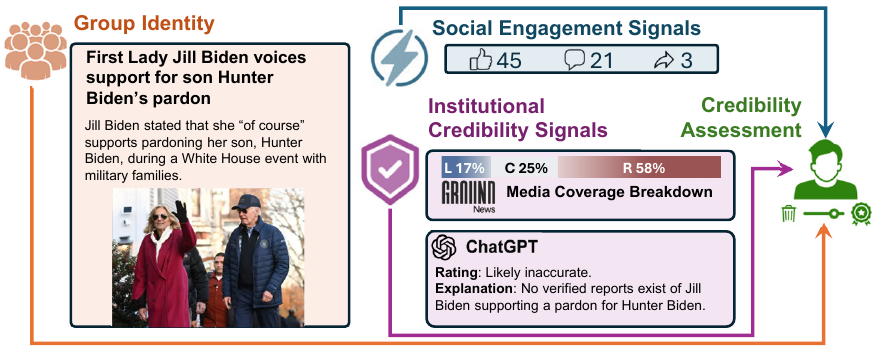}
    \caption{\textit{A sociotechnical model of credibility perception.} News consumers evaluate headline credibility based on the combined influence of social identity (ingroup alignment), engagement signals (\eg, likes, shares), and credibility cues from institutional or AI-generated sources.}
    \label{fig:teaser}
\end{figure}

A large body of research has shown that social signals, particularly those aligned with group identity, powerfully influence how users perceive and share political content~\cite{hogg2016social, huddy2001social, knobloch2015political, osmundsen2021partisan}. Ingroup bias and institutional distrust are well-documented in both online and offline settings, with individuals more likely to trust ideologically congruent sources and discount opposing views~\cite{Facciani2024, mena2020cleaning, pretus2023role}. These patterns are reinforced by platform algorithms that personalize exposure and amplify existing attitudes~\cite{epstein2015search, swart2021experiencing}, often producing self-reinforcing news environments~\cite{cinelli2021echo, grinberg2019fake}.

Against this backdrop, a new type of epistemic authority is emerging: generative AI systems that provide automated summaries, fact-checks, or credibility scores. While these signals are often perceived as neutral~\cite{binns2018s, ghani2022impact}, recent work suggests they may encode political bias~\cite{rozado2023political, hartmann2024political} or conflict with users’ existing beliefs~\cite{fujimoto2023revisiting, jia2024journalistic}. Some studies suggest that algorithmic credibility indicators can improve discernment under time pressure~\cite{pennycook2021shifting}, but others caution that overly assertive or opaque feedback may undermine trust~\cite{yin2019understanding, binns2018s}. Moreover, it remains unclear how AI-generated judgments interact with familiar social cues, such as popularity metrics or institutional labels~\cite{nah2024trust, chae2024perceiving}.

This study investigates how AI-generated credibility assessments affect users’ evaluations of politically salient news content, and how these signals interact with institutional credibility markers, social engagement metrics, and group identity. Building on prior research in social cognition, political psychology, and human-AI interaction, we test four directional hypotheses:

\begin{figure*}[t]
    \centering
    \includegraphics[width=0.99\linewidth]{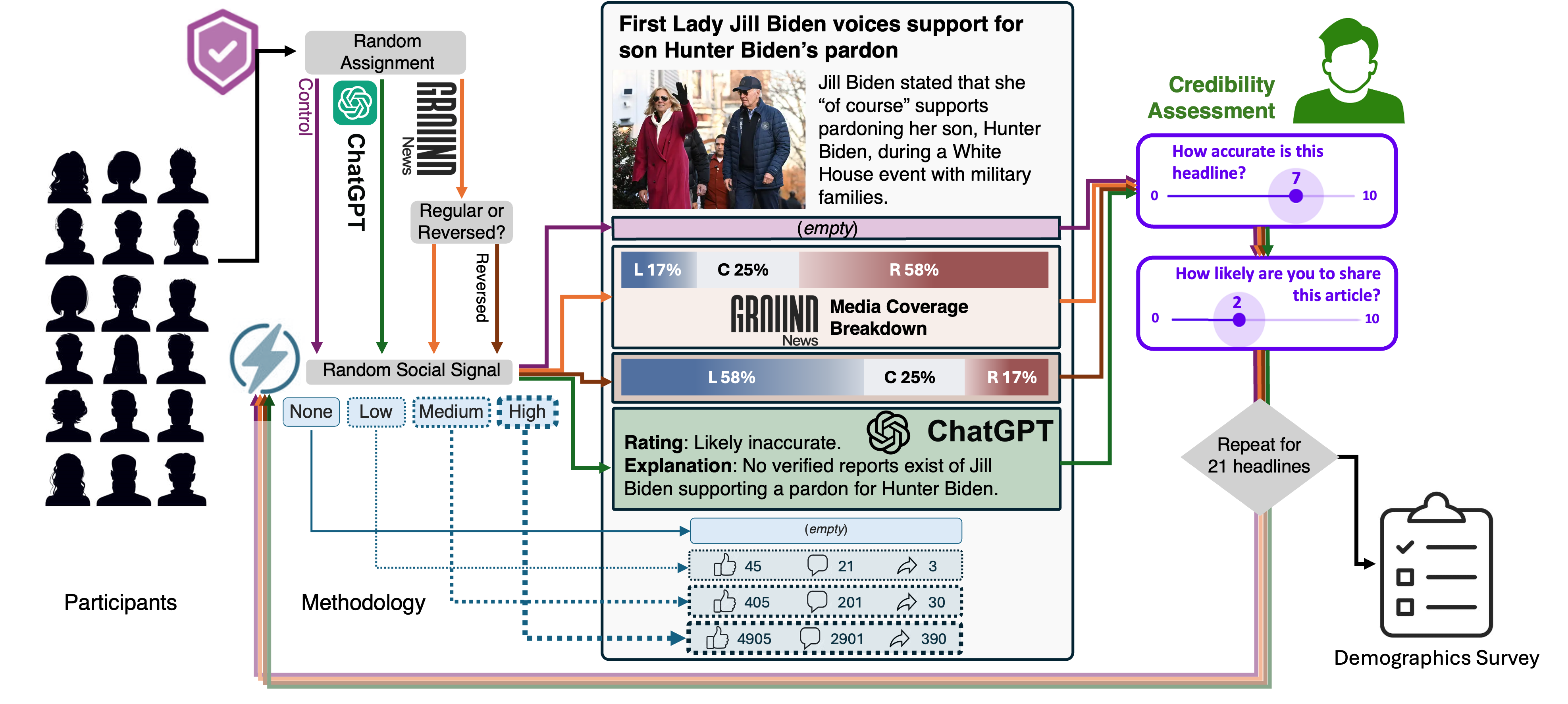}
    \caption{Overview of experimental design. Participants were randomly assigned to one of four feedback conditions: Control, GroundNews, GroundNews Reversed, or ChatGPT Feedback. For each headline shown, social engagement cues (likes, comments, shares) were drawn at random from low, medium, or high signal levels. After viewing each headline, participants rated its credibility and shareability. The session concluded with a demographics questionnaire.}
    \label{fig:method}
\end{figure*}

\begin{enumerate}[
    labelindent=1em,     
    leftmargin=2em,       
    labelwidth=2em,       
    labelsep=0em,         
    align=left
]
    \item[H1] \textbf{Ingroup Bias Hypothesis:} News headlines that align with a participant’s political affiliation will be rated as more accurate than those that oppose their views.
    
    \item[H2] \textbf{Institutional Distrust Hypothesis:} AI-generated credibility assessments will exert a greater influence on perceived accuracy than assessments from institutional sources (\eg, media bias ratings).

    \item[H3] \textbf{Social Proof Hypothesis:} Headlines accompanied by high social engagement metrics (\eg, likes, shares) will be rated as more credible than those with low or no engagement.

    \item[H4] \textbf{AI Persuasion Hypothesis:} Participants will adjust their credibility and sharing intentions in the direction of the AI-generated label (\eg, \textit{accurate} or \textit{inaccurate}), reflecting the persuasive authority of generative AI.
\end{enumerate}

Together, these hypotheses probe the interplay between identity, algorithmic authority, and social heuristics in shaping news credibility judgments. By systematically comparing traditional signals of trustworthiness with AI-generated feedback, this work offers insight into the growing influence of large language models in the sociotechnical ecosystem of online news.

\paragraph{Findings in Brief} Overall, we find that (\textbf{H1}) political identity plays a central role in shaping credibility judgments, with participants consistently rating ideologically aligned content as more accurate. (\textbf{H2}) Both AI-generated feedback and institutional signals significantly influenced these judgments, though their impact varied depending on the user’s political orientation. (\textbf{H3}) In contrast, social engagement metrics such as likes and shares had little measurable effect on perceived accuracy or shareability. (\textbf{H4}) Finally, AI-generated feedback from ChatGPT exerted the strongest influence overall, especially when it affirmed or contradicted participants’ political expectations.

\section{AI Credibility Experiment}

\subsection{Experiment Design}

We conducted a preregistered mixed-design experiment\footnote{https://osf.io/wdrpt/?view\_only=\allowbreak06cf20a0ea2241ae90a80486ef06ec3b} with one between-subjects factor (\textit{credibility feedback}) and one within-subjects factor (\textit{social engagement level}). All materials and data are publicly available\footnote{https://osf.io/kwur7/files/osfstorage?view\_only=\allowbreak4ecb53af78eb429282c474e650cb7197}.

Participants ($N = 1{,}000$) were randomly assigned to one of four feedback conditions:
\begin{enumerate}
    \item \textbf{Control}: No credibility signal.
    \item \textbf{GroundNews}: Real-world media bias classification.
    \item \textbf{Reversed GroundNews}: Inverted bias labels as a mismatch control.
    \item \textbf{ChatGPT Feedback}: AI-generated assessment of credibility.
\end{enumerate}

Each participant rated 21 political news headlines, each shown with a randomized social engagement level (none, low, medium, or high). Engagement metrics—likes, shares, comments—simulated common platform cues. Headline order and engagement assignment were randomized within subjects.

To examine group identity effects, headlines were categorized as left, center, or right-leaning using GroundNews labels. Political ingroup/outgroup alignment was computed based on participants’ self-reported affiliation.

\subsection{Participants}
Participants were recruited via Prolific with eligibility criteria including English fluency and regular news consumption. Block randomization ensured balanced demographic representation across conditions, including political affiliation, gender, race, age, education, and geography (urban/suburban/rural).

\subsection{Data}
Participants rated each headline on:
\begin{itemize}
    \item \textbf{Perceived accuracy} (0–10 scale)
    \item \textbf{Sharing intention} (0–10 scale)
\end{itemize}

The resulting dataset supports multi-level analysis and includes:
\begin{enumerate}
    \item \textbf{User data}: Demographics and media habits
    \item \textbf{Article data}: Text and ideological classification
    \item \textbf{Condition data}: Assigned feedback and engagement levels
    \item \textbf{Response data}: Accuracy and sharing judgments
\end{enumerate}

This design allows us to isolate the influence of AI feedback, social cues, and political identity on credibility assessments in a realistic, ecologically valid setting.

\subsection{Variables}

To model how sociotechnical cues influence credibility perception, we operationalize three core constructs: \textit{social identity} (ingroup bias), \textit{social proof} (engagement cues), and \textit{algorithmic authority} (credibility signals). These are captured through the following independent and dependent variables.

\subsubsection{Independent Variables (IVs)}

\begin{description}
    \item[Ingroup Alignment:] Categorical variable indicating whether the political leaning of the headline matches (\textit{ingroup}), opposes (\textit{outgroup}), or is neutral relative to the participant’s self-reported political affiliation.
    
    \item[Engagement Level:] Ordinal variable with four levels (\textit{none}, \textit{low}, \textit{medium}, \textit{high}) indicating the number of likes, shares, and comments shown.

    \item[Feedback Condition:] Between-subjects factor with four levels: \textit{Control}, \textit{GroundNews}, \textit{Reversed GroundNews}, and \textit{ChatGPT Feedback}.
    
    \item[Headline Leaning:] Categorical variable labeling each headline as \textit{left}, \textit{center}, or \textit{right}, based on GroundNews metadata.
\end{description}

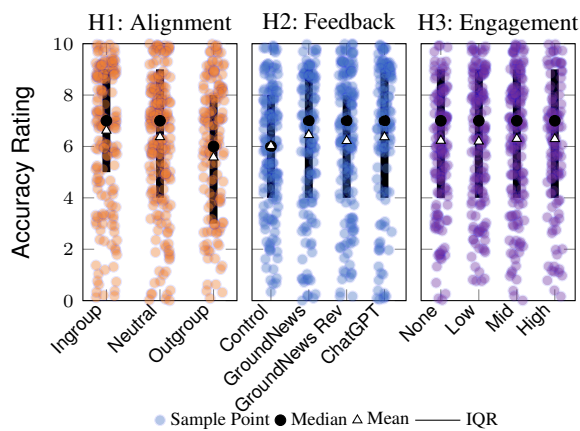
\begin{figure}[t]
    \centering
    \begin{tikzpicture}
\begin{groupplot}[
    group style={
        group size=3 by 1,
        horizontal sep=0.2cm,
        ylabels at=edge left,
    },
    width=0.45\linewidth,
    height=5cm,
    ymin=0, ymax=10,
    ylabel={Accuracy Rating},
    xlabel={},
    legend style={
        at={(0.5,-0.39)},
        anchor=north,
        draw=none,
        /tikz/every even column/.style={anchor=west},
        fill=none,
        legend columns=-1,
        font=\scriptsize,
    },
]

\nextgroupplot[
    title={H1: Alignment},
    xtick={1,2,3},
    title style={font=\small, yshift=-1.4ex},
    xticklabel style={rotate=45, anchor=east},
    xticklabels={Ingroup, Neutral, Outgroup}
]

\addplot+[
    only marks,
    mark=*,
    mark options={fill=pptorange, draw opacity=0.1,  fill opacity=0.4}
] table [x=x, y=y, col sep=comma] {h1_swarm.csv};

\addplot+[
    only marks,
    mark=*,
    mark options={fill=black, draw=black},
    forget plot
] table [x=x, y=median, col sep=comma] {h1_iqr.csv};

\addplot+[
    only marks,
    mark=triangle*,
    mark size = 2pt,
    mark options={draw=black, fill=white, solid},
    forget plot
] table [x=x, y=mean, col sep=comma] {h1_summary.csv};

\addplot+[
    mark=none,
    draw=none,
    line width=4pt,
    forget plot,
    error bars/.cd,
        y dir=plus,
        y explicit,
        error mark=bar,
    error bar style={line width=3pt, draw=black},
    error mark options={line width=3pt, draw=pptorange},            
] table [
    x=x,
    y=q1,
    y error expr=\thisrow{q3}-\thisrow{q1},
    col sep=comma
] {h1_iqr.csv};

\nextgroupplot[
    title={H2: Feedback},
    title style={font=\small, yshift=-0.9ex},
    xtick={1,2,3,4},
    xmin=0.5,
    yticklabels={},
    xticklabel style={rotate=45, anchor=east},
    xticklabels={Control, GroundNews, GroundNews Rev, ChatGPT}
]

\addplot+[
    only marks,
    mark=*,
    mark options={fill=pptblue,  draw opacity=0.1,  fill opacity=0.4}
] coordinates {(0,0)};
\addlegendentry{Sample Point}

\addplot+[
    only marks,
    mark=*,
    mark options={fill=black, draw=black, draw opacity=1,  fill opacity=1}
] coordinates {(0,0)};
\addlegendentry{Median}

\addplot+[
    only marks,
    mark=triangle*,
    mark size = 2pt,
    mark options={draw=black, fill=white, solid},
] coordinates {(0,0)};
\addlegendentry{Mean}

\addplot+[
    mark=none,
    draw=none,
    error bars/.cd,
        y dir=plus,
        y explicit,
        error mark=bar,
    error bar style={line width=3pt, draw=pptblue}
] coordinates {(0,0)};
\addlegendentry{IQR}

\addplot+[
    only marks,
    mark=*,
    mark options={fill=pptblue, draw opacity=0.1,  fill opacity=0.4}
] table [x=x, y=y, col sep=comma] {h2_swarm.csv};

\addplot+[
    only marks,
    mark=*,
    mark options={fill=black, draw=black},
    forget plot
] table [x=x, y=median, col sep=comma] {h2_iqr.csv};

\addplot+[
    only marks,
    mark=triangle*,
    mark size = 2pt,
    mark options={draw=black, fill=white, solid},
    forget plot
] table [x=x, y=mean, col sep=comma] {h2_summary.csv};

\addplot+[
    mark=none,
    draw=none,
    line width=4pt,
    forget plot,
    error bars/.cd,
        y dir=plus,
        y explicit,
        error mark=bar,
    error bar style={line width=3pt, draw=black, solid},
    error mark options={line width=3pt, draw=pptblue, solid},            
] table [
    x=x,
    y=q1,
    y error expr=\thisrow{q3}-\thisrow{q1},
    col sep=comma
] {h2_iqr.csv};

\nextgroupplot[
    title={H3: Engagement},
    title style={font=\small, yshift=-1.4ex},
    xtick={1,2,3,4},    
    yticklabels={},
    xticklabel style={rotate=45, anchor=east},
    xticklabels={None, Low, Mid, High}
]
\addplot+[
    only marks,
    mark=*,
    mark options={fill=pptpurple, draw opacity=0.1, fill opacity=0.4}
] table [x=x, y=y, col sep=comma] {h3_swarm.csv};

\addplot+[
    only marks,
    mark=*,
    mark options={fill=black, draw=black},
    forget plot
] table [x=x, y=median, col sep=comma] {h3_iqr.csv};

\addplot+[
    only marks,
    mark=triangle*,
    mark size = 2pt,
    mark options={draw=black, fill=white, solid},
    forget plot
] table [x=x, y=mean, col sep=comma] {h3_summary.csv};

\addplot+[
    mark=none,
    draw=none,
    line width=4pt,
    forget plot,
    error bars/.cd,
        y dir=plus,
        y explicit,
        error mark=bar,
    error bar style={line width=3pt, draw=black},
    error mark options={line width=3pt, draw=pptpurple},            
] table [
    x=x,
    y=q1,
    y error expr=\thisrow{q3}-\thisrow{q1},
    col sep=comma
] {h3_iqr.csv};
\end{groupplot}

\end{tikzpicture}
    \caption{Beeswarm plots with summary statistics for H1-3. Each point represents an user-input accuracy rating.}
    \label{fig:boxplot}
\end{figure}

\subsubsection{Dependent Variable (DV)}

\begin{description}
    \item[Perceived Accuracy:] Continuous rating on a 0–10 scale, indicating how accurate the participant judged the headline to be.
    
    \item[Perceived Shareability:] Continuous rating on a 0–10 scale, indicating how likely the participant would be to share the headline.
\end{description}

These variables allow us to test how identity, social cues, and algorithmic feedback interact in shaping judgments of political news. The following section presents mixed-effects regression models that quantify these relationships.

\section{Results}

We report results corresponding to the four core hypotheses. First, we test whether political alignment predicts perceived accuracy (H1: Ingroup Bias). Next, we assess the influence of institutional and AI-generated credibility signals (H2 and H4), and whether social engagement metrics shape accuracy judgments (H3).

Figure~\ref{fig:boxplot} visualizes the distribution of accuracy ratings across all experimental conditions.

\subsection{Finding 1: Political Identity Shapes Credibility Judgments}

We find strong support for H1: participants rated headlines as more accurate when the headline’s political leaning matched their own affiliation ($\beta = 2.13$, $p < 0.01$). However, this effect varied across groups. Interaction terms reveal that ingroup bias was strongest among moderates, but reversed or neutralized among liberals ($\beta = -2.14$, $p < 0.01$) and conservatives ($\beta = -2.03$, $p < 0.01$).

A two-way ANOVA confirmed a significant main effect of headline alignment ($F(2, 21097) = 37.27$, $p < 0.001$) and a strong interaction with political affiliation ($F(4, 21097) = 278.03$, $p < 0.001$). Post hoc Tukey tests showed that moderates rated ingroup-aligned content significantly higher than neutral or outgroup content ($p < 0.001$), while liberals and conservatives rated neutral headlines highest overall. These patterns suggest that ingroup effects are not symmetric across the political spectrum.

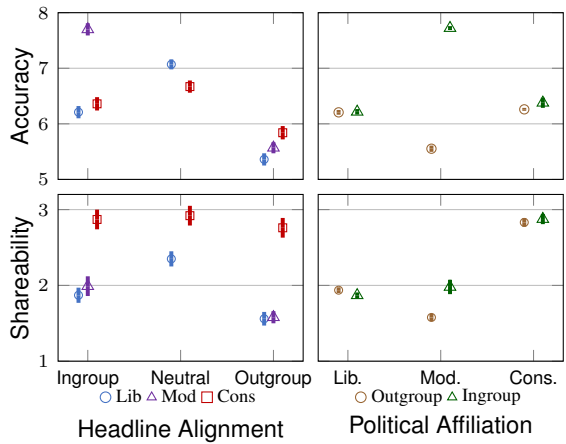
\begin{figure}[t]
    \centering
\pgfplotstableread{
alignment mean ci_lower ci_upper
ingroup 6.21 6.10 6.32
neutral 7.07 6.98 7.16
outgroup 5.36 5.25 5.47
}\datatableliberal

\pgfplotstableread{
alignment mean ci_lower ci_upper
ingroup 7.70 7.59 7.81
neutral nan nan nan
outgroup 5.57 5.47 5.67
}\datatablemoderate

\pgfplotstableread{
alignment mean ci_lower ci_upper
ingroup 6.36 6.24 6.48
neutral 6.67 6.57 6.78
outgroup 5.84 5.73 5.96
}\datatableconservative

\pgfplotstableread{
alignment mean ci_lower ci_upper
ingroup 1.87 1.77 1.97
neutral 2.35 2.24 2.45
outgroup 1.56 1.47 1.65
}\sharedatatableliberal

\pgfplotstableread{
alignment mean ci_lower ci_upper
ingroup 1.99 1.86 2.12
neutral nan nan nan
outgroup 1.58 1.49 1.66
}\sharedatatablemoderate

\pgfplotstableread{
alignment mean ci_lower ci_upper
ingroup 2.87 2.75 3.00
neutral 2.92 2.79 3.05
outgroup 2.76 2.63 2.89
}\sharedatatableconservative

\pgfplotstableread{
term    coef   ci_low  ci_high
Moderate	5.569379	5.402428	5.73633
Conservative 0.689997	0.471057	0.908937	
Liberal	0.645753	0.436017	0.85548
Ingroup~(Mod)	2.129098	1.977822	2.280375	
Ingroup~×~Cons  -2.029014	-2.227398	-1.830630
Ingroup~×~Lib -2.135023	-2.325066	-1.944979	
Group~Var	0.189566	0.168286	0.210846
}\coefplotdata

\pgfplotstableread{
term    coef   ci_low  ci_high
Moderate                2.84    2.61    3.07
Conservative           -1.27   -1.62   -0.91
Liberal                -0.89   -1.20   -0.57
Ingroup~(Mod)           0.03   -0.06    0.13
Ingroup~×~Cons          0.38    0.23    0.53
Ingroup~×~Lib          -0.12   -0.25    0.01
}\sharecoefdata

\pgfplotstableread{
is_ingroup	political_affiliation	predicted	ci_lower	ci_upper
0	Liberal	6.206578	6.146023	6.249339
0	Moderate	5.552454	5.513878	5.608546
0	Conservative	6.260565	6.231998	6.286778
}\partialdataoutgroup

\pgfplotstableread{
is_ingroup	political_affiliation	predicted	ci_lower	ci_upper
1	Liberal	6.216659	6.160932	6.265871
1	Moderate	7.722724	7.688197	7.757807
1	Conservative	6.378952	6.318660	6.470033
}\partialdataingroup

\pgfplotstableread{
is_ingroup	political_affiliation	predicted	ci_lower	ci_upper
0    Liberal    1.9369007059855339    1.9113177700632273    1.974801018805114
0    Moderate    1.5771190152471823    1.5313364225896562    1.6237362119964958
0    Conservative    2.8312285964134483    2.803518073732658    2.8885322545412166
}\sharepartialdataoutgroup

\pgfplotstableread{
is_ingroup	political_affiliation	predicted	ci_lower	ci_upper
1    Liberal    1.869558412392292    1.8197594456012638    1.9071542960121655
1    Moderate    1.9794792302966417    1.886192807727524    2.0742519031731033
1    Conservative    2.8752916192542903    2.7832150272089353    2.9422855864350907
}\sharepartialdataingroup

\begin{tikzpicture}
\begin{groupplot}[
    group style={
        group size=2 by 2,
        horizontal sep=0.2cm,
        vertical sep=0.2cm,
    },    
    legend style={
        at={(0.5,-0.1)},
        anchor=north,
        draw=none,
        /tikz/every even column/.style={anchor=west},
        fill=none,
        font=\scriptsize,
    },
    ymajorgrids=true,
    width=0.6\linewidth,
    height=3.8cm,
]

\nextgroupplot[
    xtick={1,2,3},
    ylabel={Accuracy},
    ylabel style={yshift=-5.0pt},
    ymin=5, ymax=8,
    xticklabels={},
    xlabel={},
    legend columns=3
]

\addplot+[
    only marks,
    mark=o,
    mark options={fill=pptblue, draw=pptblue},
    mark size=1.5pt,
    forget plot,
    error bars/.cd,
        y dir=both,
        y explicit,
        error mark=bar,
        error bar style={line width=1.5pt, draw=pptblue},
] table[
    x expr=\coordindex + 0.9,
    y=mean,
    y error expr=\thisrow{ci_upper} - \thisrow{mean}
] \datatableliberal;

\addplot+[
    only marks,
    mark=triangle,
    mark options={fill=pptpurple, draw=pptpurple},
    mark size=2.5pt,
    forget plot,
    error bars/.cd,
        y dir=both,
        y explicit,
        error mark=bar,
        error bar style={line width=1.5pt, draw=pptpurple},
] table[
    x expr=\coordindex + 1,
    y=mean,
    y error expr=\thisrow{ci_upper} - \thisrow{mean}
] \datatablemoderate;

\addplot+[
    only marks,
    mark=square,
    mark options={fill=pptred, draw=pptred},
    mark size=1.5pt,
    forget plot,
    error bars/.cd,
        y dir=both,
        y explicit,
        error mark=bar,
        error bar style={line width=1.5pt, draw=pptred},
] table[
    x expr=\coordindex + 1.1,
    y=mean,
    y error expr=\thisrow{ci_upper} - \thisrow{mean}
] \datatableconservative;

\nextgroupplot[
    xtick={1,2,3},
    ylabel={},    
    yticklabels={},
    ymin=5, ymax=8,
    xticklabels={},
    xlabel={},
    legend columns=2
]

\addplot+[
    only marks,
    mark=o,
    mark options={fill=pptbrown, draw=pptbrown},
    mark size=1.75pt,
    forget plot,
    error bars/.cd,
        y dir=both,
        y explicit,
        error mark=bar,
        error bar style={line width=1.5pt, draw=pptbrown},
] table[
    x expr=\coordindex + 0.9,
    y=predicted,
    y error expr=\thisrow{ci_upper} - \thisrow{predicted}
] \partialdataoutgroup;

\addplot+[
    only marks,
    mark=triangle,
    mark options={fill=pptgreen, draw=pptgreen},
    mark size=2.5pt,
    forget plot,
    error bars/.cd,
        y dir=both,
        y explicit,
        error mark=bar,
        error bar style={line width=1.5pt, draw=pptgreen},
] table[
    x expr=\coordindex + 1.1,
    y=predicted,
    y error expr=\thisrow{ci_upper} - \thisrow{predicted}
] \partialdataingroup;

\nextgroupplot[
    xtick={1,2,3},
    ylabel={Shareability},
    ylabel style={yshift=-5.0pt},
    xlabel={Headline Alignment},
    xlabel style={yshift=-5.0pt},
    ymin=1, ymax=3.2,
    ytick={1,2,3},
    xticklabels={Ingroup, Neutral, Outgroup},
    legend columns=3,
]

\addplot+[
    only marks,
    mark=o,
    mark options={fill=pptblue, draw=pptblue},
    mark size=1.5pt,
    forget plot,
    error bars/.cd,
        y dir=both,
        y explicit,
        error mark=bar,
        error bar style={line width=1.5pt, draw=pptblue},
] table[
    x expr=\coordindex + 0.9,
    y=mean,
    y error expr=\thisrow{ci_upper} - \thisrow{mean}
] \sharedatatableliberal;
\addlegendimage{only marks, mark=o, mark options={fill=pptblue, draw=pptblue}}
\addlegendentry{Lib}

\addplot+[
    only marks,
    mark=triangle,
    mark options={fill=pptpurple, draw=pptpurple},
    mark size=2.5pt,
    forget plot,
    error bars/.cd,
        y dir=both,
        y explicit,
        error mark=bar,
        error bar style={line width=1.5pt, draw=pptpurple},
] table[
    x expr=\coordindex + 1,
    y=mean,
    y error expr=\thisrow{ci_upper} - \thisrow{mean}
] \sharedatatablemoderate;
\addlegendimage{only marks, mark=triangle, mark options={fill=pptpurple, draw=pptpurple}}
\addlegendentry{Mod}

\addplot+[
    only marks,
    mark=square,
    mark options={fill=pptred, draw=pptred},
    mark size=1.5pt,
    forget plot,
    error bars/.cd,
        y dir=both,
        y explicit,
        error mark=bar,
        error bar style={line width=1.5pt, draw=pptred},
] table[
    x expr=\coordindex + 1.1,
    y=mean,
    y error expr=\thisrow{ci_upper} - \thisrow{mean}
] \sharedatatableconservative;
\addlegendimage{only marks, mark=square, mark options={fill=pptred, draw=pptred}}
\addlegendentry{Cons}

\nextgroupplot[
    xtick={1,2,3},
    ylabel={},    
    yticklabels={},
    ymin=1, ymax=3.2,
    ytick={1,2,3},
    xticklabels={Lib., Mod., Cons.},
    xlabel={Political Affiliation},
    xlabel style={yshift=-5.0pt},
    legend columns=2
]

\addplot+[
    only marks,
    mark=o,
    mark options={fill=pptbrown, draw=pptbrown},
    mark size=1.5pt,
    forget plot,
    error bars/.cd,
        y dir=both,
        y explicit,
        error mark=bar,
        error bar style={line width=1.5pt, draw=pptbrown},
] table[
    x expr=\coordindex + 0.9,
    y=predicted,
    y error expr=\thisrow{ci_upper} - \thisrow{predicted}
] \sharepartialdataoutgroup;
\addlegendimage{only marks, mark=o, mark options={fill=pptbrown, draw=pptbrown}}
\addlegendentry{Outgroup}

\addplot+[
    only marks,
    mark=triangle,
    mark options={fill=pptgreen, draw=pptgreen},
    mark size=2.5pt,
    forget plot,
    error bars/.cd,
        y dir=both,
        y explicit,
        error mark=bar,
        error bar style={line width=1.5pt, draw=pptgreen},
] table[
    x expr=\coordindex + 1.1,
    y=predicted,
    y error expr=\thisrow{ci_upper} - \thisrow{predicted}
] \sharepartialdataingroup;
\addlegendimage{only marks, mark=triangle, mark options={fill=pptgreen, draw=pptgreen}}
\addlegendentry{Ingroup}

\end{groupplot}
\end{tikzpicture}
    \caption{
Predicted accuracy (top) and shareability (bottom) ratings by headline alignment (left) and political affiliation (right). Points show group means with 95\% confidence intervals. Ingroup effects are strongest among moderates, while partisans show a consistent preference for center-aligned content and limited ingroup bias in sharing.
    }
    \label{fig:h1-results}
\end{figure}

Figure~\ref{fig:h1-results} (top row) shows this divergence. Moderates show a clear ingroup preference; partisans show a preference for center-aligned content over both ingroup and outgroup headlines. On average, neutral headlines received higher accuracy ratings than left-leaning ($\beta = 0.594$, $p < 0.001$) and right-leaning ($\beta = 0.420$, $p < 0.001$) headlines, consistent with prior findings~\cite{hartmann2024political}.

\subsection*{Ingroup Bias and Shareability}

While accuracy ratings showed clear evidence of ingroup effects, we next examine whether these biases extend to behavioral intentions, specifically participants' judgments about how shareable each article was. The pattern diverges meaningfully.

The mixed-effects model results reveal a modest but selective ingroup effect on shareability. Only political moderates show a statistically significant ingroup boost ($\beta = 0.38$, $p < 0.01$), suggesting they are more inclined to share content that aligns with their political identity (\ie, neutrality). In contrast, liberals rated ingroup content as slightly less shareable than neutral content, and conservatives showed no significant difference.

A two-way ANOVA further confirms this conditionality. There was no significant main effect of alignment on shareability ratings ($F(2, 21097) = 1.13$, $p = 0.29$), but there was a strong main effect of political affiliation ($F(2, 21097) = 186.50$, $p < 0.001$), as well as a robust interaction between alignment and affiliation ($F(4, 21097) = 26.47$, $p < 0.001$). These results suggest that political identity alone does not shape sharing intentions unless it interacts with headline alignment.

Tukey post hoc tests reveal that moderates were significantly more likely to share ingroup (\ie, neutral) content than outgroup content ($p < 0.001$), while liberals and conservatives showed no such preference. In some cases, they even rated ingroup content as less shareable than center-aligned headlines. These patterns mirror but attenuate those observed for credibility ratings, implying that credibility is a necessary but insufficient condition for sharing.

In sum, ideological alignment influences both belief formation and willingness to share, but the effect on behavior is less uniform. Moderates appear uniquely sensitive to alignment in both cases. For other groups, shareability judgments seem to reflect a blend of credibility, identity signaling, and a broader preference for moderate content.

\subsection{Finding 2: Institutional vs. Algorithmic Credibility Signals}

\begin{figure}
    \centering
    \pgfplotstableread{
idx x         coef  ci_low  ci_high               label          type
0  0.9  6.024666  5.950765  6.098567             Control      Accuracy
2  1.9  6.450216  6.369145  6.531288     GroundNews      Accuracy
4  2.9  6.217634  6.140623  6.294645       GroundNews~Rev      Accuracy
6  3.9  6.365730  6.286848  6.444612  ChatGPT      Accuracy
}\dataplotacc

\pgfplotstableread{
idx x         coef  ci_low  ci_high               label          type
1  1.1  2.040767  1.965880  2.115655             Control  Shareability
3  2.1  1.942692  1.860539  2.024845     GroundNews  Shareability
5  3.1  2.252232  2.174193  2.330271       GroundNews~Rev  Shareability
7  4.1  2.534738  2.454804  2.614673  ChatGPT  Shareability
}\dataplotshare

\begin{tikzpicture}
\begin{groupplot}[
    group style={
        group size=2 by 1,
        horizontal sep=1.5cm,
    }, 
    height=5.3cm,
    width=4.5cm,
    ymajorgrids=true,
    legend style={
        at={(0.5,-0.15)},
        anchor=north,
        draw=none,
        /tikz/every even column/.style={anchor=west},
        fill=none,
        font=\scriptsize,
    },
    ymajorgrids=true,
]
\nextgroupplot[
    ylabel={Accuracy Rating},
    xlabel={},
    xtick=data,
    xticklabels from table={\dataplotacc}{label},
    xticklabel style={rotate=45, anchor=east},
    enlargelimits=0.10,
    ymajorgrids=true,
    ymin=5.5, ymax=7.05,
    ytick={5, 6, 7},
    legend style={
        at={(1,1)},
        anchor=north east,
        draw=none,
        /tikz/every even column/.style={anchor=west},
        fill=none,
        font=\scriptsize,
    },
    xmajorgrids=true,
]

\addplot+[
    only marks,
    mark=o,
    mark size = 2.3pt,
    mark options={fill=pptdarkgrey, draw=pptdarkgrey},
    error bars/.cd,
        y dir=both,
        y explicit,
        error mark=bar,
        error bar style={line width=2.5pt, draw=pptdarkgrey},
] table[
    y=coef,
    x expr=\coordindex,
    y error plus expr=\thisrow{ci_high} - \thisrow{coef},
    y error minus expr=\thisrow{coef} - \thisrow{ci_low}
] \dataplotacc;

\draw[gray] (0,5.8) -- (1,5.8);  
\draw[gray] (0,5.8) -- (0,5.85);  
\draw[gray] (1,5.8) -- (1,5.85);  
\node[gray, font=\footnotesize] at (0.5,5.8) {***};

\draw[gray] (0,5.7) -- (2,5.7);
\draw[gray] (0,5.7) -- (0,5.75);
\draw[gray] (2,5.7) -- (2,5.75);
\node[gray,font=\footnotesize] at (1,5.7) {**};

\draw[gray] (0,5.6) -- (3,5.6);
\draw[gray] (0,5.6) -- (0,5.65);
\draw[gray] (3,5.6) -- (3,5.65);
\node[gray,font=\footnotesize] at (1.5,5.6) {***};

\draw[gray] (1,6.7) -- (2,6.7);
\draw[gray] (1,6.7) -- (1,6.65);
\draw[gray] (2,6.7) -- (2,6.65);
\node[gray,font=\footnotesize] at (1.5,6.7) {***};

\draw[gray] (2,6.6) -- (3,6.6);
\draw[gray] (2,6.6) -- (2,6.55);
\draw[gray] (3,6.6) -- (3,6.55);
\node[gray,font=\footnotesize] at (2.5,6.6) {*};

\draw[gray] (1,6.8) -- (3,6.8);
\draw[gray] (1,6.8) -- (1,6.75);
\draw[gray] (3,6.8) -- (3,6.75);
\node[gray,font=\footnotesize] at (2,6.8) {};

\nextgroupplot[
    ylabel={Shareability Rating},
    xlabel={},
    xtick=data,
    xticklabels from table={\dataplotacc}{label},
    xticklabel style={rotate=45, anchor=east},
    enlargelimits=0.10,
    ymajorgrids=true,
    ymin=1.5, ymax=3.05,
    ytick={1, 2, 3},
    legend style={
        at={(1,1)},
        anchor=north east,
        draw=none,
        /tikz/every even column/.style={anchor=west},
        fill=none,
        font=\scriptsize,
    },
    xmajorgrids=true,
]

\addplot+[
    only marks,
    mark=square,
    mark options={fill=pptdarkgreen, draw=pptdarkgreen},
    mark size=1.85pt,
    error bars/.cd,
        y dir=both,
        y explicit,
        error mark=bar,
        error bar style={line width=2.5pt, draw=pptdarkgreen},
] table[
    y=coef,
    x expr=\coordindex,
    y error plus expr=\thisrow{ci_high} - \thisrow{coef},
    y error minus expr=\thisrow{coef} - \thisrow{ci_low}
] \dataplotshare;

\draw[gray] (0,1.8) -- (1,1.8);
\draw[gray] (0,1.8) -- (0,1.85);
\draw[gray] (1,1.8) -- (1,1.85);
\node[gray, font=\footnotesize] at (0.5,1.8) {};

\draw[gray] (0,1.7) -- (2,1.7);
\draw[gray] (0,1.7) -- (0,1.75);
\draw[gray] (2,1.7) -- (2,1.75);
\node[gray, font=\footnotesize] at (1,1.7) {***};

\draw[gray] (0,1.6) -- (3,1.6);
\draw[gray] (0,1.6) -- (0,1.65);
\draw[gray] (3,1.6) -- (3,1.65);
\node[gray, font=\footnotesize] at (1.5,1.6) {***};

\draw[gray] (1,2.7) -- (2,2.7);
\draw[gray] (1,2.7) -- (1,2.65);
\draw[gray] (2,2.7) -- (2,2.65);
\node[gray, font=\footnotesize] at (1.5,2.7) {***};

\draw[gray] (2,2.6) -- (3,2.6);
\draw[gray] (2,2.6) -- (2,2.55);
\draw[gray] (3,2.6) -- (3,2.55);
\node[gray, font=\footnotesize] at (2.5,2.6) {***};

\draw[gray] (1,2.8) -- (3,2.8);
\draw[gray] (1,2.8) -- (1,2.75);
\draw[gray] (3,2.8) -- (3,2.75);
\node[gray, font=\footnotesize] at (2,2.8) {***};

\end{groupplot}
\end{tikzpicture}
    \caption{
    Mean Accuracy and Shareability ratings by feedback condition. Left: Estimated accuracy ratings with 95\% confidence intervals. Right: Corresponding shareability ratings. Stars denote significant differences from Tukey’s HSD tests (* $p < .05$, ** $p < .01$, *** $p < .001$). All feedback increased perceived accuracy over the control, with GroundNews producing the strongest effect. ChatGPT generated the largest increase in shareability.
    }
    \label{fig:h2_means}
\end{figure}
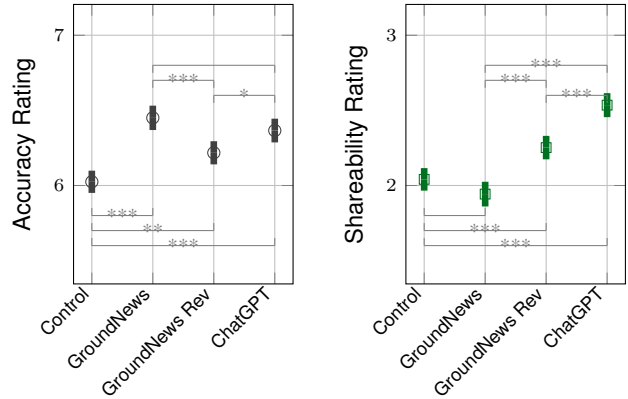

We next tested H2: the Institutional Distrust Hypothesis, which posits that users may perceive AI-generated credibility assessments (\eg, ChatGPT) as more persuasive than traditional institutional cues (\eg, GroundNews), particularly when political alignment differs. This reflects a broader question about whether algorithmic feedback is viewed as more neutral or credible in polarized media environments.

\subsubsection{Accuracy Judgments}

As shown in Fig.~\ref{fig:h2_means} (left), all credibility signals increased perceived accuracy relative to the control. GroundNews produced the strongest effect, followed closely by ChatGPT, partially supporting H2. While institutional cues remain influential, algorithmic feedback emerged as a comparably powerful source of epistemic authority.

\begin{figure}
    \centering
    \input{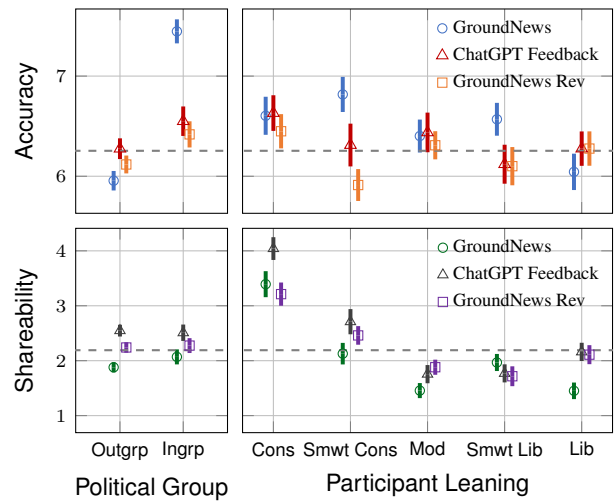}
\caption{
Effects of credibility signals on accuracy and shareability.
Left: Predicted ratings by headline alignment under GroundNews and Reversed GroundNews conditions. 
Right: Predicted ratings across political affiliations under all feedback conditions. 
Error bars show 95\% confidence intervals (Tukey post hoc). Dotted lines indicate the control mean. 
Alignment boosts credibility under GroundNews but weakens or reverses under Reversed. 
ChatGPT produces consistent ratings across groups and is notably more persuasive among conservatives.
}
    \label{fig:h2_mlm_marginals}
\end{figure}

A two-way ANOVA revealed main effects of both feedback condition ($F(2, 15345) = 8.84$, $p < .001$) and headline alignment ($F(1, 15345) = 186.17$, $p < .001$), as well as a strong interaction ($F(2, 15345) = 65.29$, $p < .001$). As illustrated in Fig.~\ref{fig:h2_mlm_marginals} (top-left), GroundNews was highly effective when aligned with the user’s ideology but was discounted when misaligned. ChatGPT feedback, in contrast, boosted accuracy ratings consistently across alignment conditions, suggesting higher perceived neutrality.

A second model examined feedback by political affiliation ($F(8, 15273) = 6.37$, $p < .001$). GroundNews had the strongest effect among liberals and moderates but was significantly less persuasive for conservatives. ChatGPT elicited relatively uniform accuracy ratings across all ideological groups (Fig.~\ref{fig:h2_mlm_marginals}, top-right), further supporting the hypothesis that algorithmic feedback may avoid partisan asymmetries in trust.

\subsubsection{Shareability Judgments}

ChatGPT feedback also increased willingness to share headlines. In a two-way ANOVA, feedback condition significantly affected shareability ($F(1, 10223) = 30.49$, $p < .001$), while ideological alignment had no effect. As shown in Fig.~\ref{fig:h2_mlm_marginals} (bottom-left), ChatGPT increased shareability across the board; institutional cues produced smaller and less consistent changes.

Feedback by political affiliation yielded similar results ($F(8, 15273) = 9.22$, $p < .001$): ChatGPT feedback improved shareability across groups, while GroundNews was less effective and, for conservatives, even counterproductive (Fig.~\ref{fig:h2_means}, right).

\subsubsection{Summary and Interpretation of H2}
Together, these results strongly support H2. Institutional credibility signals were filtered through political identity—effective when aligned, dismissed when not. In contrast, ChatGPT feedback was broadly persuasive, increasing both accuracy and shareability regardless of user ideology.

Additional mixed-effects models (not shown) confirm that ChatGPT had the largest overall influence: accuracy ratings increased when it labeled content as credible and decreased when it flagged content as inaccurate. GroundNews, by contrast, produced a uniform credibility boost, suggesting less content-sensitive impact. These results indicate that algorithmic feedback—when framed with clear explanation—may function as a more adaptable and inclusive credibility intervention in politically diverse environments.

\subsection{Finding 3: Limited Influence of Social Engagement Metrics}

Next we tested (H3) The Social Engagement Hypothesis: that visible engagement cues (likes, shares, comments) act as social proof, increasing perceived credibility and shareability of news headlines. Engagement levels were varied across four conditions: none, low, medium, and high.

\subsubsection{Accuracy Judgments}
A one-way ANOVA showed no significant effect of engagement level on perceived accuracy ($F(3, 21185) = 1.60$, $p = .19$). A mixed-effects model with participant-level random effects confirmed this result: engagement cues had no reliable impact on credibility ratings. These findings suggest that passive social signals do not meaningfully shape perceived accuracy in this setting.

\subsubsection{Shareability Judgments}
Shareability ratings showed a weak trend ($F(3, 21185) = 2.37$, $p = .069$), but post hoc tests found no significant pairwise differences ($p > .11$). A mixed-effects model again found no significant effect of engagement level on sharing behavior.

\subsubsection{Summary and Interpretation of H3}

Contrary to H3, engagement metrics had no significant effect on credibility or shareability. Unlike algorithmic or institutional signals, these cues lacked persuasive power in our controlled setting. One possible explanation is context: stripped from familiar platform environments, engagement metrics may appear arbitrary or uninformative. These results suggest that the influence of social proof may depend heavily on context, salience, and user familiarity—important considerations for designing credibility interventions outside native platforms.

\begin{figure}
    \centering
    \input{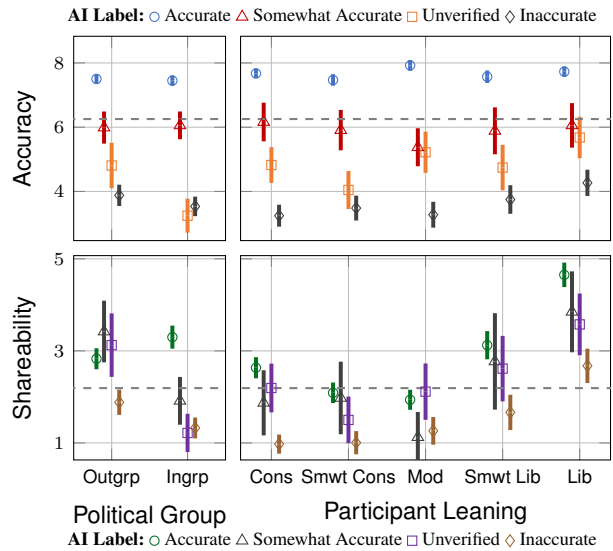}
    \caption{Effects of ChatGPT feedback on perceived accuracy (top) and shareability (bottom). Left: Mean ratings by feedback type across political affiliations. 
Right: Ratings by political affiliation across feedback types. 
Headlines labeled \textit{Accurate} received the highest ratings; those labeled \textit{Inaccurate}, the lowest. 
Liberals responded most strongly to negative feedback, while conservatives showed more moderate shifts.
}
    \label{fig:h4tukey}
\end{figure}

\subsection{Finding 4: AI Feedback Strongly Shapes Credibility Judgments}

We tested the AI Feedback Hypothesis (H4): that participants would adjust credibility and shareability ratings in response to ChatGPT-generated labels (\textit{accurate}, \textit{somewhat accurate}, \textit{unverified}, \textit{inaccurate}). This hypothesis reflects the growing epistemic authority of generative AI in digital media environments.

\subsubsection{Accuracy Judgments}

A two-way ANOVA revealed a strong effect of feedback type on perceived accuracy ($F(3, 5062) = 831.62$, $p < .001$), with ratings scaling predictably across the four labels. There was also a main effect of political leaning ($F(4, 5062) = 5.98$, $p < .001$) and a significant interaction between feedback and political identity ($F(12, 5062) = 3.49$, $p < .001$), suggesting that trust in AI feedback is filtered through users’ ideological priors.

\subsubsection{Interaction with Ideological Alignment}

A separate ANOVA confirmed that political alignment significantly moderated the effect of ChatGPT feedback ($F(6, 5070) = 12.08$, $p < .001$). As shown in Fig.~\ref{fig:h4tukey} (top), ratings increased when ChatGPT labeled ingroup headlines as \textit{accurate}, and sharply declined when it labeled them \textit{inaccurate}. This asymmetry illustrates how algorithmic judgments amplify or dampen trust depending on users' prior beliefs.

\subsubsection*{Effect on Sharing Behavior}

Shareability ratings followed a similar pattern. ChatGPT feedback boosted sharing intentions overall, especially for ingroup content. As shown in Fig.~\ref{fig:h4tukey} (bottom), participants were more likely to share headlines labeled \textit{accurate} when those headlines aligned with their political identity. The effect was strongest under ambiguous labels like \textit{somewhat accurate}, where ingroup headlines were nearly twice as likely to be shared as outgroup.

\subsubsection{Summary of H4 Findings}

These results strongly support H4. Participants adjusted both credibility and sharing judgments in response to ChatGPT’s feedback—particularly when the content aligned with their political identity. While this suggests that AI feedback can nudge users away from low-quality content, it also raises concerns about overreliance. When algorithmic judgments reinforce existing biases, users may reduce critical scrutiny, deferring too readily to AI authority.
Designing AI interventions that are persuasive yet preserve user agency remains a key challenge. These findings underscore the need for transparency, explainability, and careful framing when deploying credibility signals in real-world systems.

\begin{figure}[t]
    \centering
    \input{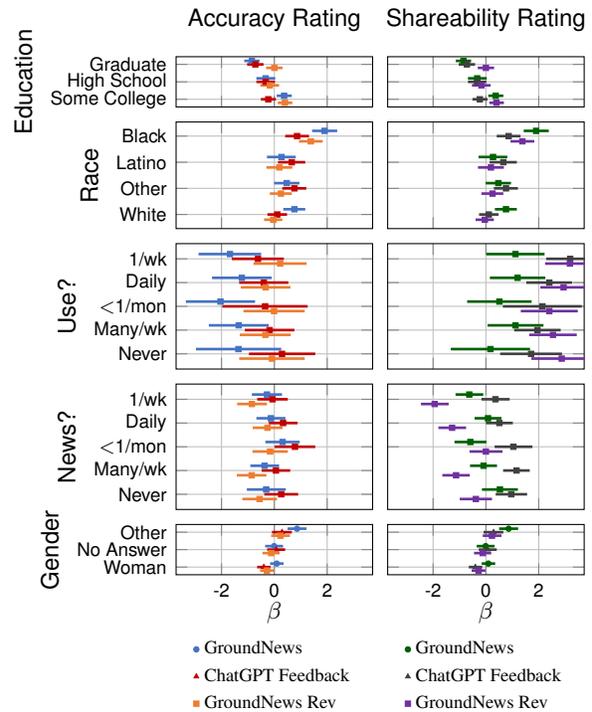}
\caption{
Interaction effects of demographics and feedback condition on accuracy and shareability.
Each subplot shows OLS coefficients for interactions between feedback type and a demographic variable. 
Error bars represent 95\% confidence intervals. 
ChatGPT feedback had the strongest overall impact on accuracy; effects on sharing were more variable. 
}
    \label{fig:explore_coef}
\end{figure}

\section{Exploratory Analysis}

To examine how user demographics moderate the effects of credibility signals, we fit ordinary least squares (OLS) models with main effects and interaction terms for feedback condition, education, race, gender, social media use, and news consumption frequency.

\subsubsection{Accuracy Judgments}

ChatGPT feedback significantly increased perceived accuracy overall, but this effect was weaker among highly educated users and frequent social media consumers, who expressed more skepticism. In contrast, GroundNews and Reversed labels reduced perceived accuracy among those same groups, suggesting that overt bias markers may backfire among more media-literate participants. Women and racially minoritized participants (particularly Black and Latine users) responded more positively to AI-based feedback than institutional labels.

\subsubsection{Shareability Judgments}

GroundNews feedback, especially when reversed, produced the sharpest declines in sharing intentions, particularly among frequent news and social media users. ChatGPT’s effects were more variable: it reduced sharing for negatively labeled headlines but increased shareability among several subgroups, including Black and Latino participants. Graduate degree holders showed heightened responsiveness to all feedback types across both outcomes.

\subsubsection*{Summary of Findings}

Figure~\ref{fig:explore_coef} summarizes interaction effects by demographic group. These patterns underscore the importance of tailoring credibility interventions to diverse audiences and suggest that demographic context meaningfully shapes how users interpret AI and institutional signals.

\section{Discussion}

This study examined how political identity, social cues, and feedback mechanisms shape perceptions of news credibility. Our results support three of four hypotheses and offer several key insights:

\begin{itemize}
    \item \textbf{H1} \textit{Political Identity}: Supported. Ideological alignment strongly influenced perceived accuracy, with moderates favoring centrist content and partisans showing ingroup bias.
    
    \item \textbf{H2} \textit{Institutional Distrust}: Partially supported. GroundNews signals increased credibility when politically aligned but were discounted otherwise. ChatGPT feedback produced more consistent effects across groups.

    \item \textbf{H3} \textit{Social Engagement}: Not supported. Likes, shares, and comments had no significant impact on accuracy or shareability, suggesting weak influence outside of native social contexts.

    \item \textbf{H4} \textit{AI Feedback}: Supported. ChatGPT feedback reliably influenced user judgments, especially when feedback contradicted prior beliefs. Effects were observed for both accuracy and sharing.
\end{itemize}

These findings highlight both the potential and risk of algorithmic feedback in shaping public understanding. AI-generated cues can help mitigate bias and enhance credibility discernment, but their influence varies by political identity and carries ethical risks related to overreliance.

\subsection{Implications}

This study examined how political identity, credibility signals, and social metrics influence users’ perceptions of news accuracy and shareability. We found that political alignment strongly shaped credibility judgments, consistent with social identity theory and selective exposure \cite{seering2018applications,liao2015all,tajfel1979integrative}. Even when presented with the same headlines, participants rated ideologically congruent content as more accurate, with moderates showing a preference for centrist headlines. These findings highlight political moderates as a particularly receptive audience for credibility interventions.

AI-generated feedback from ChatGPT significantly influenced both accuracy and shareability ratings, often overriding institutional signals and even moderating partisan effects. This suggests growing trust in algorithmic credibility assessments, though it raises concerns about over-reliance and the need for transparent, explainable AI systems.

In contrast, social engagement metrics had no meaningful effect on perceived accuracy. This may reflect the artificiality of the experimental setting or a general decline in the persuasive power of likes and shares when not embedded in authentic social contexts \cite{winter2015they}. It remains possible that the persuasive power of likes and shares depends heavily on contextual familiarity and peer networks \cite{muchnik2013social, bakshy2012role,de2012popularity}. 

\subsection{Design Considerations}

AI-driven credibility tools—when well-calibrated—can support more informed evaluations, especially when paired with visual explanations or institutional context. However, to be trusted and effective across ideological divides, such tools must be:
\begin{itemize}
    \item transparent in their criteria,
    \item adaptable to diverse users,
    \item and designed to enhance, not override, human judgment.
\end{itemize}

Hybrid approaches that combine algorithmic feedback with prompts for reflection may better preserve user agency and avoid the risks of epistemic outsourcing.

\subsection{Limitations and Future Research Directions}

This study used a controlled online setting, which may not fully capture the dynamics of real-world news consumption. Future work should explore these effects in organic social environments and across other generative models or feedback formats (e.g., tone, uncertainty, or visual cues).

We also found limited effects for engagement metrics, but future studies could test whether such cues more strongly influence sharing behavior than perceived credibility. The outsized influence of ChatGPT highlights a critical trade-off: while persuasive, AI feedback may displace critical thinking if not carefully framed.

Moving forward, research should focus on designing AI credibility interventions that are persuasive yet transparent, adaptable to user context, and supportive of epistemic autonomy—especially in politically polarized and low-trust environments.

\section{Conclusions}

This study advances our understanding of how political identity, social cues, and AI-generated credibility signals shape perceptions of news accuracy. The results highlight three key findings: (1) political identity remains a dominant influence on credibility judgments, (2) AI feedback—particularly from ChatGPT—exerts strong, cross-partisan effects on both belief and behavior, and (3) social engagement metrics play a limited role outside native platform contexts.

These findings have direct implications for the design of credibility interventions in sociotechnical systems. Users are increasingly influenced by algorithmic feedback, which can override institutional cues and moderate partisan bias—but also risks promoting overreliance. Institutional signals remain effective for some users, but their impact diminishes in politically polarized or low-trust environments. Meanwhile, engagement metrics such as likes and shares were largely ignored, suggesting reduced persuasive value when presented without social context.

To support equitable and informed news evaluation, AI-driven interventions must be transparent, explanatory, and designed to enhance user agency. Future work should examine these mechanisms in more ecologically valid settings, evaluate alternative AI credibility framings, and develop adaptive systems that foster critical engagement across politically diverse audiences.

\bibliographystyle{plainnat}
\bibliography{main}


\appendix

\end{document}